\documentclass[journal]{IEEEtran}
\ifCLASSINFOpdf
\else
   \usepackage{graphicx}
   \graphicspath{{../eps/}}
   \DeclareGraphicsExtensions{.eps}
\fi

\usepackage{amsmath}
\usepackage{amsfonts,amssymb}
\usepackage{amssymb}
\usepackage{array}
\usepackage{wrapfig}
\usepackage{psfrag}
\usepackage{epstopdf}
\usepackage{cite}
\usepackage{graphicx}
\usepackage{subfigure}
\usepackage{threeparttable}
\usepackage{cases}
\usepackage{subeqnarray}
\usepackage{color}
\usepackage{underscore}
\usepackage{verbatim}
\usepackage{bm}
\usepackage{stfloats}
\usepackage{xpatch}
\usepackage{makecell}

\usepackage{algorithm}
\usepackage{algorithmic}
\usepackage[table]{xcolor}

\begin{document}
\title{Universal CKM for Environment-Aware\\ Wireless Networks: Enabling Cross-Device\\ and Cross-Task Channel Knowledge Transfer
}
%
%
%%
%%
%% author names and IEEE memberships
%% note positions of commas and nonbreaking spaces ( ~ ) LaTeX will not break
%% a structure at a ~ so this keeps an author's name from being broken across
%% two lines.
%% use \thanks{} to gain access to the first footnote area
%% a separate \thanks must be used for each paragraph as LaTeX2e's \thanks
%% was not built to handle multiple paragraphs
%%
%
\author{Haiquan~Lu,~\IEEEmembership{Member,~IEEE,}
        Yong~Zeng,~\IEEEmembership{Fellow,~IEEE,}
        Cheng-Xiang~Wang,~\IEEEmembership{Fellow,~IEEE,}
%        Shi~Jin,~\IEEEmembership{Fellow,~IEEE,}
        Xiqi~Gao,~\IEEEmembership{Fellow,~IEEE,}
%        Jianhua~Zhang,~\IEEEmembership{Fellow,~IEEE,}
        and Rui~Zhang,~\IEEEmembership{Fellow,~IEEE}
%%        Long~Shi,~\IEEEmembership{Senior Member,~IEEE,}\\
%%        Shaodan~Ma,~\IEEEmembership{Senior Member,~IEEE,}
%%        Shi~Jin,~\IEEEmembership{Fellow,~IEEE,}
%%        and
%%        Rui~Zhang,~\IEEEmembership{Fellow,~IEEE}
%%%        % <-this % stops a space
%%\thanks{This work was supported in part by the Natural Science Foundation for Distinguished Young Scholars of Jiangsu Province under Grant BK20240070, in part by the National Natural Science Foundation of China under Grant 62071114, and in part by the Fundamental Research Funds for the Central Universities under Grant 2242022k60004. (\emph{Corresponding author: Yong Zeng.}) }
\thanks{Haiquan Lu is with the School of Electronic and Optical Engineering, Nanjing University of Science and Technology, Nanjing 210094, China (e-mail: haiquanlu@njust.edu.cn).}
\thanks{Yong Zeng, Cheng-Xiang Wang, and Xiqi Gao are with the National Mobile Communications Research Laboratory, Southeast University, Nanjing 210096, China, and also with the Purple Mountain Laboratories, Nanjing 211111, China (e-mail: \{yong_zeng, chxwang, xqgao\}@seu.edu.cn). (\emph{Corresponding author: Yong Zeng.})}
\thanks{Rui Zhang is with the Department of Electrical and Computer Engineering, National University of Singapore, Singapore 117583 (e-mail: elezhang@nus.edu.sg).}
%%%
%%% <-this % stops a space
}

\maketitle

 \begin{abstract}
 Channel knowledge map (CKM) is a promising technology for environment-aware sixth-generation (6G) wireless networks. However, existing CKMs are tightly coupled with wireless devices and downstream tasks, which limit their scalability and reusability in wireless networks. To address these limitations, this article proposes the concept of universal CKM (uCKM) as a foundational wireless environment prior, which aims to enable cross-device and cross-task channel knowledge transfer for environment-aware wireless networks. We first revisit the representative CKMs and discuss their limitations. Then, the uCKM-enabled new paradigm for environment-aware wireless networks is introduced, and its benefits are highlighted from the perspectives of uCKM construction and utilization phases, for which we propose the visions of ``All for uCKM'' and ``uCKM for All'', i.e., the data acquired by all devices and tasks should contribute to the construction of uCKM, while the constructed uCKM, in turn, can be utilized to support all devices and tasks. Subsequently, we discuss the main challenges of uCKM and propose potential solutions. Last, we provide simulation results to demonstrate the feasibility and performance gains brought by uCKM and outline future research directions. 
 \end{abstract}

%\begin{IEEEkeywords}
% %Low-altitude economy, UAV swarm, near-field, aerial movable antenna (AMA), swarm trajectory optimization.
%\end{IEEEkeywords}

\IEEEpeerreviewmaketitle
% >>>>>>>>>>>>>SECTIONS I -  here >>>>>>>>>>>>
\section{Introduction}
\begin{figure}[!t]
 \centering
 \centerline{\includegraphics[width=3.5in,height=2.1in]{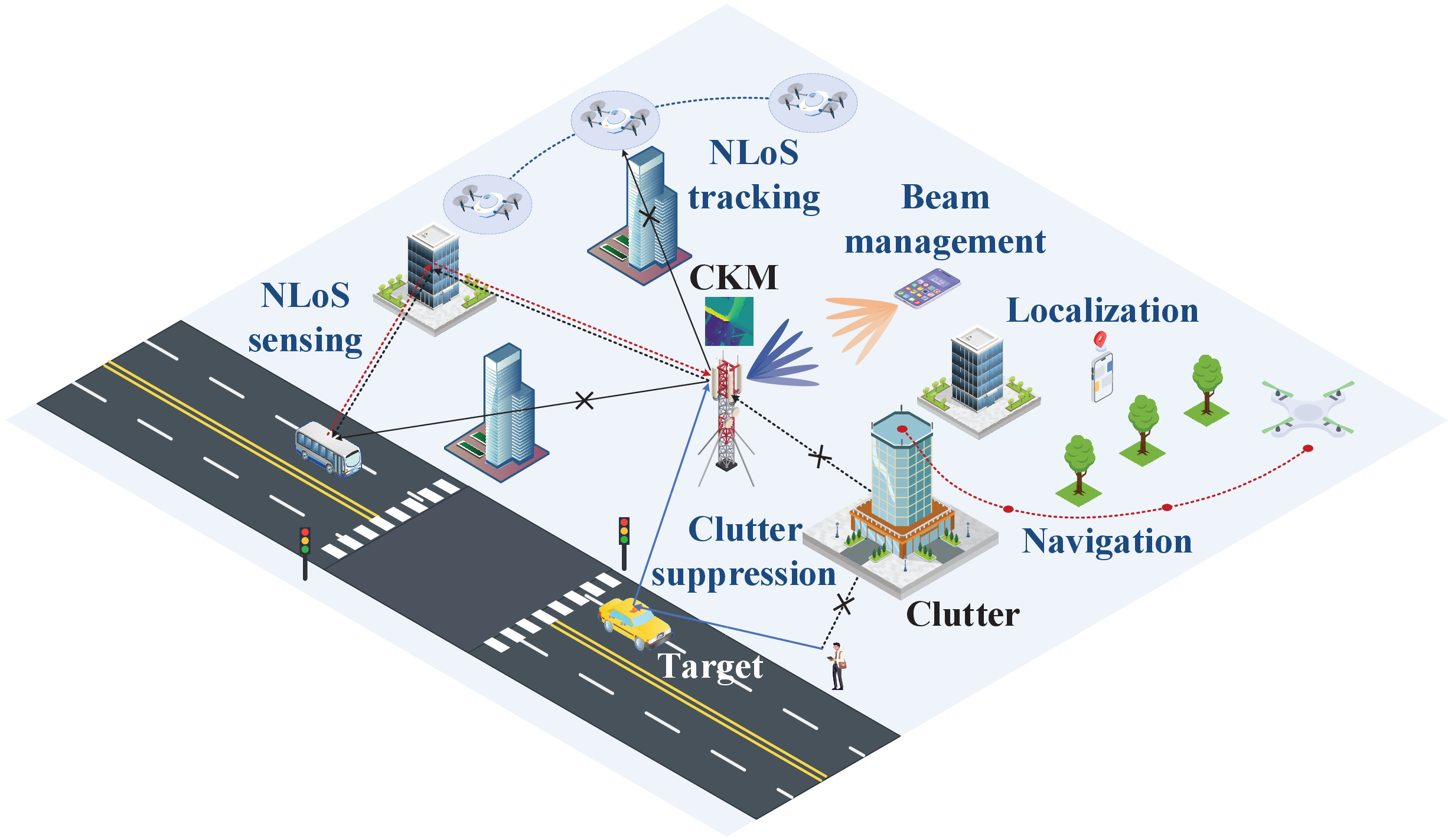}}
 \caption{Illustration of CKM applications, including communication task such as beam management; localization task such as NLoS localization and geometry-based localization; sensing task such as NLoS sensing, NLoS tracking, and clutter suppression; as well as navigation task.}
 \label{fig:CKMApplications}
 \end{figure}

 Channel knowledge map (CKM) has emerged as a promising technology for enabling environment-aware wireless networks \cite{zeng2021toward}. As a site-specific digital database or artificial intelligence (AI) model, CKM tries to learn from the massive location-tagged data and establish a mapping from the transceiver location to channel prior knowledge, such as line-of-sight (LoS)/non-LoS (NLoS) states, channel gain, angle-of-arrival/departure (AoA/AoD), and propagation delay. With such channel knowledge priors, diverse downstream tasks, including communication, sensing, localization, and navigation, can operate in an environment-aware manner \cite{zeng2021toward,liu2025channel}. This avoids repeatedly performing a prior channel estimation or environment sensing from scratch, which is particularly appealing for sixth-generation (6G) networks characterized by large spatial and frequency dimensions of wireless channels and denser node deployments \cite{zeng2024tutorial,he2024direct}. Building upon this capability, CKM has found extensive applications in wireless networks, as illustrated in Fig.~\ref{fig:CKMApplications}. For example, CKM can facilitate environment-aware network planning, resource allocation, beam management, and interference mitigation \cite{wang2025radio,liu2025channel}. Beyond wireless communications, CKM has been also used for localization, sensing and clutter suppression in complex NLoS environment \cite{wu2026you}. Moreover, with the rapid development of the low-altitude economy, CKM can provide the three-dimensional (3D) environmental priors for LoS/NLoS link identification, trajectory planning and aerial navigation for unmanned aerial vehicles (UAVs).
 
 On the other hand, wireless communication, localization, and sensing tasks usually require different forms of location-dependent channel knowledge. For instance, for wireless communications, the large-scale channel gain is usually sufficient for network planning to determine the deployment strategy of base stations (BSs). For beam management, the set of most probable candidate beam indices are required to perform fast beam selection. In contrast, localization and sensing usually benefit from multi-path characteristics that capture the geometric features of the surrounding environment, such as AoA/AoD and propagation delay. To accommodate these diverse knowledge requirements, different types of CKMs have been proposed, such as channel map (CM) \cite{qi2026data}, channel gain map (CGM) \cite{xu2024much,sun2026channel}, beam index map (BIM) \cite{zeng2024tutorial}, and channel path map (CPM) \cite{jiang2025interference}. %Besides, earlier works have mainly focused on techniques such as radio map \cite{bi2019engineering}, which typically characterizes the signal power spectrum density. 
 In the following, four representative types of CKMs are discussed.

 \subsubsection{CM}
 CM directly learns the complete location-specific time-domain channel impulse response, whose dimension increases with the number of transmit and receive antennas, as well as that of the channel taps. As an alternative, the location-specific frequency-domain transfer function can be learned. 
 %For efficient storage, the transfer function can be discretized with the channel coherence bandwidth. By exploiting the fact that channel coherence bandwidth is typically inversely proportional to the channel delay spread, the number of sub-bands can be approximated by that of channel taps. Thus, the dimension of the frequency-domain channel knowledge is equal to its time-domain counterpart. 
 Despite preserving the most complete channel knowledge, conventional CM suffers from practical issues, including the substantial storage overhead and the difficulty of reliably associating instantaneous channel coefficients with location. To alleviate these issues, learning-based methods, such as neural radio-frequency radiance field (NeRF$^2$), neural wireless radiation field (NeWRF), WRF-Gaussian splatting (WRF-GS), WRF-GS+, and F$^4$-CKM, have been proposed to learn the WRF representations, rather than explicitly storing the complete channel impulse response \cite{zhou2026f}. However, these learning-based methods usually require task-specific environment or RF measurement information as input for channel state information (CSI) prediction.

 \subsubsection{CGM}
 Compared to CM, CGM is usually simpler since it only focuses on the channel strength. For wideband frequency-selective channel, the dimension of channel knowledge is equal to the number of channel taps, which is independent of the number of transmit and receive antennas. Besides, one common practice is to store the average value across all channel taps or sub-bands, and thus the dimension of channel knowledge is reduced to one. However, CGM inevitably discards fine-grained channel information, which limits its application in downstream tasks.

 \subsubsection{BIM}
 For massive multiple-input multiple-output (MIMO) and extremely large-scale MIMO (XL-MIMO) systems, BIM provides an efficient approach by directly learning the location-specific optimal beam pair or a small set of candidate beam pairs, thus bypassing the explicit high-dimensional channel representation and enabling a fast beam management.  %Since BIM stores the indices of candidate transmit and receive beamforming vectors, the dimension of channel knowledge is $J_{\rm BIM} = 2{B_t}{B_r}$, where $B_t$ and $B_r$ denote the numbers of candidate transmit and receive beam indices, respectively. %In particular, the dimension decreases to $J_{\rm BIM} = 2$ when only the optimal beam index is stored at the transmitter and receiver. 
 
 \subsubsection{CPM} 
 CPM aims to learn the location-specific multi-path parameters, including the complex-valued path gain, elevation and azimuth AoDs, AoAs, and the propagation delay. Since the complex-valued path gain involves both real and imaginary parts, the dimension of channel knowledge is $7L$, where $L$ denotes the number of multi-paths. Note that CPM is appealing for massive MIMO and XL-MIMO communications over high-frequency channels with multi-path sparsity, such as millimeter wave (mmWave) and Terahertz channels. In this case, given the transceiver characteristics, CPM can in principle reconstruct the complete channel impulse response with a much smaller dimension than CM. However, existing CPMs are typically defined in the local coordinate systems that are severely coupled with the antenna configuration and orientation of the associated devices, and the complex-valued path gain is sensitive to location variations and localization errors.

 Despite their effectiveness in respective application scenarios, existing CKMs have two major limitations: 1) incapable of cross-device CKM sharing: specifically, CM, CGM, BIM, and CPM are severely coupled with device-specific transceiver configurations, such as antenna number, array geometry, device orientation, and beam codebook. For example, the CGM learned for one device (say a mobile phone) is useless for another device (say a vehicle), even when they operate in exactly the same environment. This greatly limits the scalability and usability of CKM; 2) limited cross-task CKM sharing: existing CKMs are usually tailored to specific downstream tasks, resulting in fragmented knowledge representations. For example, BIM stores the location-specific candidate beam indices for communication-oriented beam management. Although these indices may convey coarse direction information, they cannot explicitly provide the detailed environment and path information required by sensing tasks, such as AoA/AoD, scatterer and obstacle locations, as well as clutter characteristics. Thus, such task-specific designs lead to isolated CKM silos and limited cross-task knowledge sharing. %{\color{red}It is also worth mentioning that the recent learning-based methods have enabled the cross-task reuse of learned WRF representations, whereas a unified task-decoupled wireless propagation knowledge representation remains unexplored.}

  \begin{figure}
 \centering
 \subfigure[Channel knowledge transfer across heterogeneous devices]{
 \begin{minipage}[t]{0.5\textwidth}
 \centering
 \centerline{\includegraphics[width=2.5in,height=1.1in]{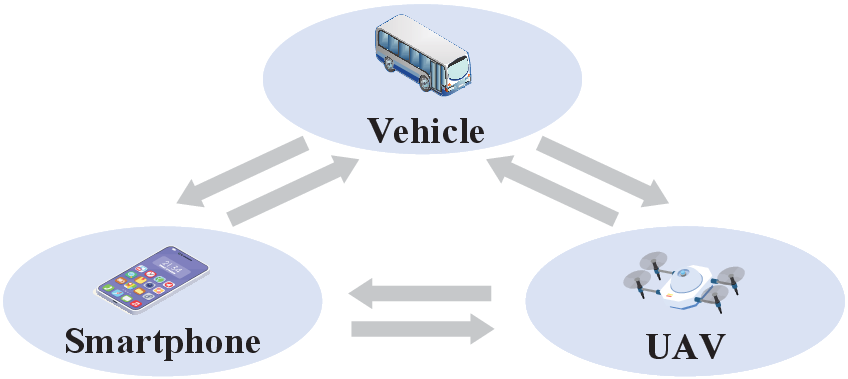}}
 \end{minipage}
 }
 \subfigure[Channel knowledge transfer across diverse downstream tasks]{
 \begin{minipage}[t]{0.5\textwidth}
 \centering
 \centerline{\includegraphics[width=2.5in,height=1.1in]{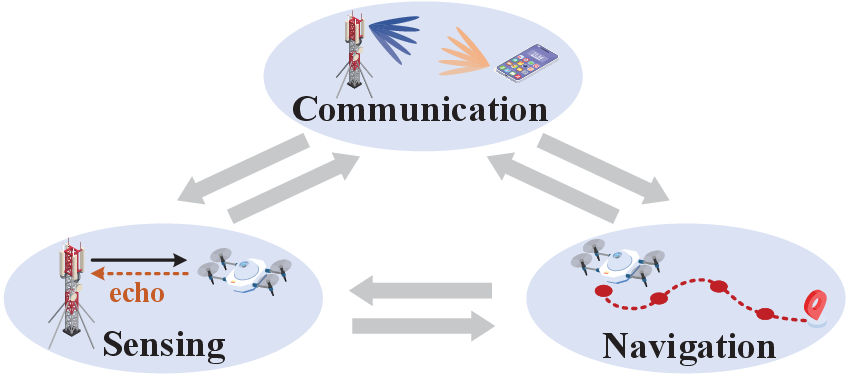}}
 \end{minipage}
 }
 \subfigure[uCKM as a channel knowledge prior base for all devices and tasks]{
 \begin{minipage}[t]{0.5\textwidth}
 \centering
 \centerline{\includegraphics[width=3.5in,height=1.4in]{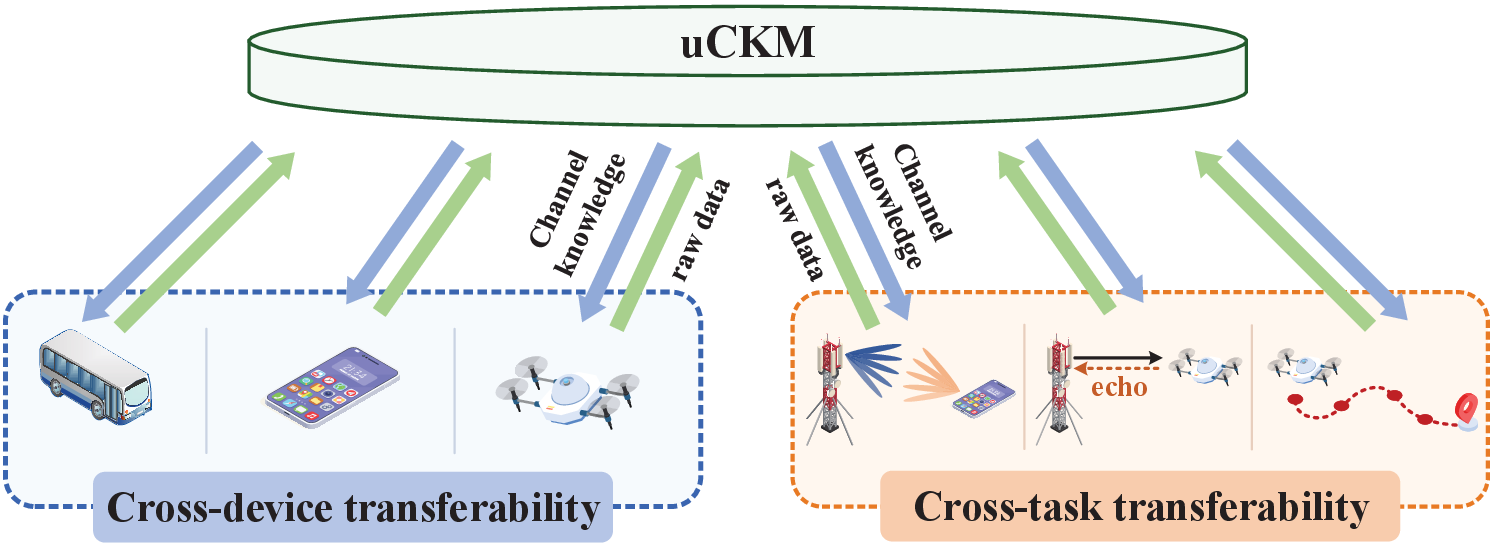}}
 \end{minipage}
 }
 \caption{uCKM-enabled cross-device and cross-task knowledge transfer.}
 \label{fig:uCKMChannelKnowledgeTransfer}
 \end{figure}
 
 The above limitations naturally raise a fundamental question: \emph{As illustrated in Fig.~\ref{fig:uCKMChannelKnowledgeTransfer}, is it possible to construct a unified CKM that enables channel knowledge transfer across all devices and downstream tasks?} This question motivates the concept of universal CKM (uCKM). Specifically, uCKM is a foundational wireless environment prior that learns underlying propagation knowledge independent of specific devices or tasks, thus allowing a single uCKM to be shared and reused across heterogeneous devices and downstream tasks within a given area of interest. In this context, the term ``universal'' refers to such cross-device and cross-task reusability, rather than the universality across arbitrary environments. If uCKM is feasible, then for each area of interest, only a single uCKM needs to be constructed, which is essential to unlock the full potential of CKM and facilitate its practical implementation. 
 
\begin{figure}[!t]
 \centering
 \centerline{\includegraphics[width=3.5in,height=2.5in]{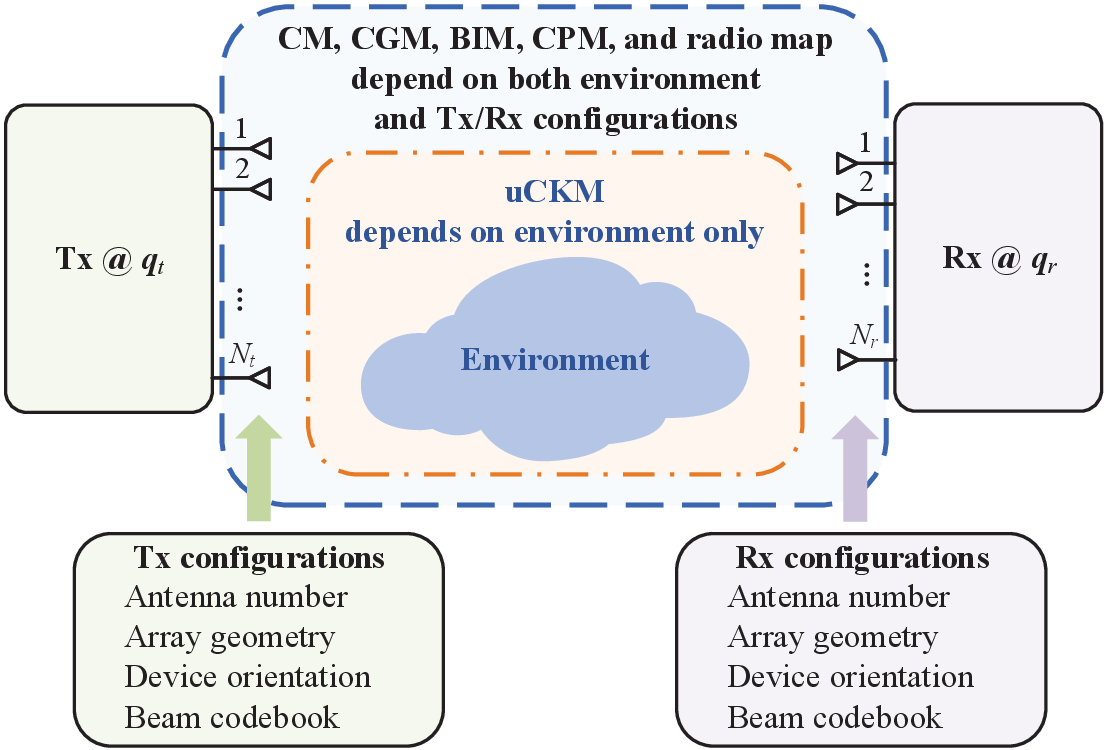}}
 \caption{Illustration of uCKM, which depends on the wireless propagation environment only, while independent of the Tx/Rx configurations or operations.}
 \label{fig:differentPartsOfChannel}
 \end{figure}

 To achieve the above vision, uCKM needs to satisfy two essential requirements: 
 \begin{itemize}[\IEEEsetlabelwidth{12)}]
 \item \textbf{Underlying wireless propagation knowledge:} uCKM should only learn the underlying wireless propagation knowledge that is governed by the surrounding environment, while independent of the device-specific characteristics such as the transmitter/receiver (Tx/Rx) antenna configurations, as shown in Fig.~\ref{fig:differentPartsOfChannel}. Such knowledge could be those multi-path parameters, including AoA/AoD, path amplitude, and propagation delay. 
 \item \textbf{Unified global coordinate system and probabilistic representation:} uCKM should be defined based on a unified global coordinate system, rather than in the local coordinate systems as in conventional channel representation approaches. Moreover, to accommodate environmental dynamics, the mapping from the transceiver location ${\bf q}$ to channel knowledge ${\bf z}$ can be represented as a location-conditional probability density function (pdf) \cite{wu2026you}, i.e., $p\left( {\left. {\bf{z}} \right|{\bf{q}}} \right)$, rather than as a purely deterministic mapping.
 \end{itemize} 
  %, where ${\bf q}$ and ${\bf z}$ denote the transceiver location and channel knowledge, respectively,
 
 Note that in contrast to conventional CPMs that typically establish a deterministic mapping from transceiver locations to specific multi-path parameters defined in local coordinate systems, uCKM learns the location-conditional pdf of wireless propagation knowledge defined in a unified global coordinate system. Besides, for different operating frequency bands, the frequency-dependent propagation characteristics can be incorporated into ${\bf z}$ and jointly represented within the same $p\left( {\left. {\bf{z}} \right|{\bf{q}}} \right)$. Moreover, the above two key requirements jointly lay the foundation for uCKM to provide a reusable wireless environment prior. Specifically, capturing the device-decoupled propagation knowledge renders it possible to transfer the learned channel knowledge across heterogeneous devices, while representing such knowledge in a unified global coordinate system with a probabilistic form makes it alignable and fusible across different data sources, as well as robust to environmental dynamics. Then, global data collected from either channel estimation in communication tasks or path information extraction in sensing tasks is represented as a unified location-conditional pdf, and diverse downstream tasks can exploit the common channel knowledge to infer the customized knowledge according to device configurations and task requirements.
 
 In this article, we provide an overview of uCKM and discuss the uCKM-enabled paradigm from two complementary perspectives, i.e., construction and utilization phases, respectively. Then, the challenges towards realizing uCKM and potential solutions are discussed. Furthermore, the simulation results are presented to demonstrate the feasibility and performance gains of uCKM.

% >>>>>>>>>>>>>SECTIONS II -  here >>>>>>>>>>>>
 \section{uCKM-Enabled Environment-Aware Wireless Networks}
\begin{figure*}[!t]
 \centering
 \centerline{\includegraphics[width=5.5in,height=3.5in]{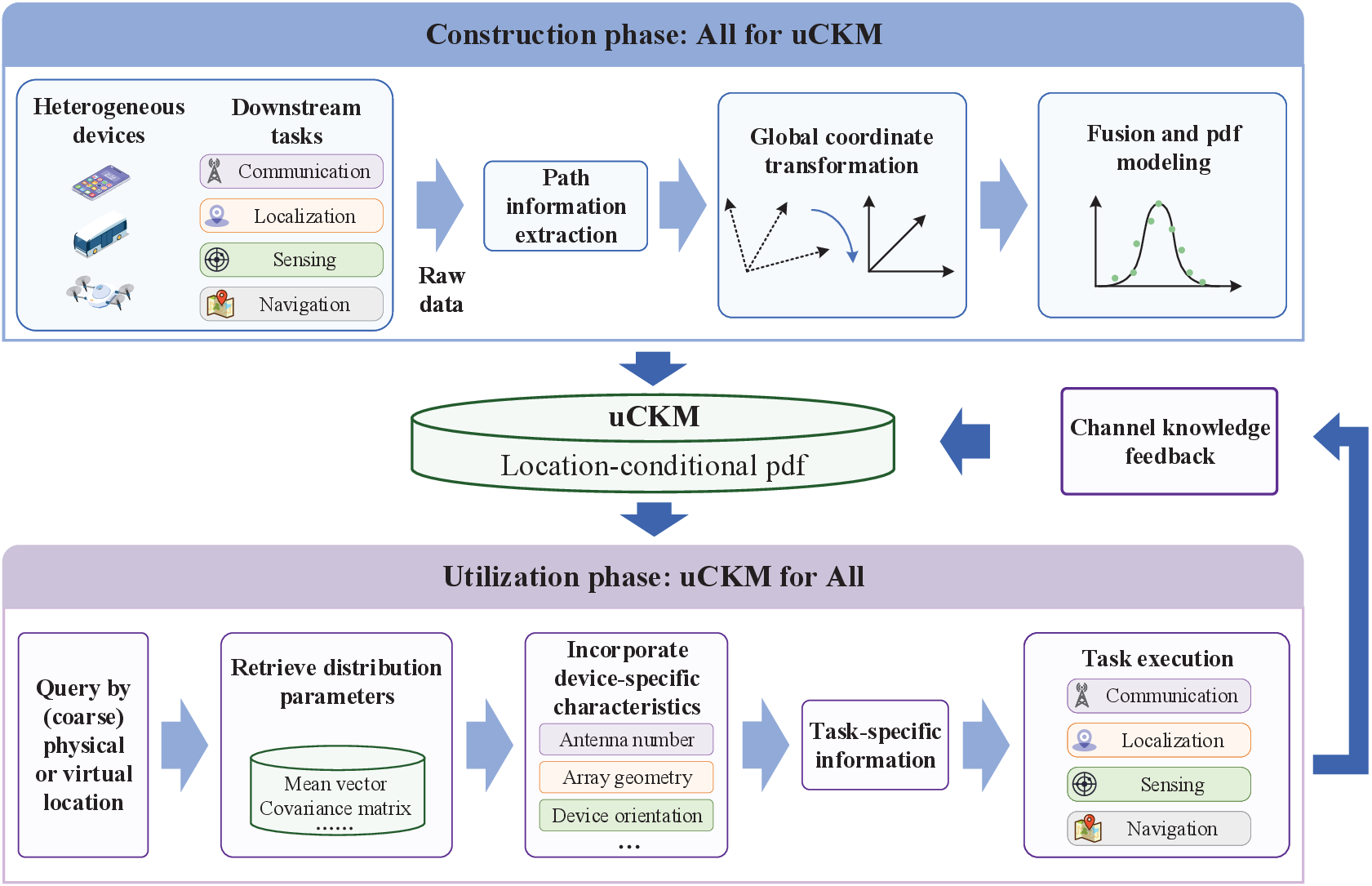}}
 \caption{The uCKM-enabled new paradigm for environment-aware wireless networks, where heterogeneous devices and downstream tasks collaboratively construct uCKM under the vision of ``All for uCKM'', and the constructed uCKM is utilized to support heterogeneous devices and tasks under the vision of ``uCKM for All''.}
 \label{fig:uCKMParadigm}
 \end{figure*}

 With the shared wireless propagation knowledge, uCKM enables a closed-loop paradigm for environment-aware wireless networks, which consists of the construction and utilization phases, as illustrated in Fig.~\ref{fig:uCKMParadigm}. This differs from existing learning-based methods that mainly focus on one-way knowledge reuse from learned WRF representations to downstream tasks, while uCKM further enables heterogeneous devices and diverse downstream tasks to contribute to the construction and updating. Specifically, in the construction phase, the data collected from heterogeneous devices and  different downstream tasks are collaboratively fused to construct uCKM, a vision which we term as ``All for uCKM''. In the utilization phase, the constructed uCKM serves as a common environment prior base that can be leveraged to support heterogeneous devices and different downstream tasks, which we refer to as ``uCKM for All''. Such a paradigm is significantly different from conventional wireless communication, localization and sensing systems, where the estimated CSI and probed environmental information are typically confined to the associated devices and tasks, rather than accumulated for cross-device and cross-task reuse. Moreover, uCKM goes beyond isolated and fragmented CKMs by unifying device- and task-decoupled propagation knowledge into a shared wireless environment prior.

 As illustrated in Fig.~\ref{fig:uCKMParadigm}, during the construction phase, the massive heterogeneous devices, such as smartphones, vehicles, UAVs, and sensing nodes, serve as the natural sensors and collaboratively contribute to a crowdsourced wireless knowledge acquisition process. To this end, each device needs to decouple its local observations from the transceiver configuration and extract the path information. With the position and orientation information of the device, the obtained path information is transformed into a unified global coordinate system. Then, the globally aligned tuples including the transceiver locations and corresponding channel knowledge are fused to construct the location-conditional pdf. The probabilistic distribution model can be selected according to the propagation scenario and statistical characteristics of channel knowledge. For example, a Gaussian distribution may be suitable for an approximately unimodal channel knowledge distribution, whereas the Gaussian mixture distribution can provide a more generic representation when multiple statistical components or clusters exist. Compared to the deterministic representation, such a probabilistic representation conveys richer information by indicating not only the most likely realization of channel knowledge, but also the associated uncertainty. Besides, the probabilistic representation can reduce the channel knowledge dimension, since only the distribution parameters are stored. For example, the Gaussian distribution can be compactly characterized by its first- and second-order statistics, i.e., mean vector and covariance matrix. As a result, uCKM converts fragmented observations from individual devices into a shared wireless propagation knowledge, and the probabilistic form captures the uncertainty caused by environmental dynamics and measurement noise. In particular, the probabilistic uCKM degenerates to the deterministic uCKM, when the conditional pdf reduces to a Dirac delta function.
  
 On the other hand, during the utilization phase, each device queries uCKM according to its (coarse) physical or virtual location and recovers the location-conditional pdf based on the distribution parameters, where the physical location corresponds to the device's actual geometric location, whereas the virtual location refers to a privacy-preserving representation of its physical location, thus enabling devices to query uCKM without directly exposing their exact physical locations. Such a privacy-preserving mapping can be achieved through existing location privacy-preserving mechanisms (LPPMs). Moreover, to ensure compatibility, the data format and generation procedure of the virtual location should be standardized in terms of the representation structure, dimensionality, associated metadata, and privacy-preserving mapping. Subsequently, by incorporating the device-specific characteristics, such as antenna number, array geometry, device orientation, and polarization modes, the device-specific channel distribution can be derived and transformed into the task-specific information. Specifically, for wireless communication, the obtained channel distribution can be used to derive the channel matrix or candidate beam indices, thus reducing the pilot or beam sweeping overhead. For localization, the location-conditional pdf can be used to evaluate the likelihood of the observed channel knowledge at each candidate location, and the device location can be obtained by selecting the location with the maximum likelihood. Moreover, the environment prior provided by uCKM can be utilized to suppress clutter, identify targets, and infer environmental changes. In particular, the construction and utilization phases form a closed-loop knowledge cycle, which transforms uCKM into a self-evolving wireless environment prior for future environment-aware networks. In the following, we elaborate on the benefits in terms of ``uCKM for All'' and ``All for uCKM''.

 \subsection{uCKM for All}

 \subsubsection{Cross-Device Channel Knowledge Transfer}% (Transferability) 
 By capturing the device-agnostic wireless propagation knowledge, uCKM provides a common environment prior that can be shared across heterogeneous devices. This enables rapid adaptation to newly deployed devices or changes in device configurations, without requiring extensive additional measurements, model retraining, or even constructing a new CKM from scratch. With the proliferation of heterogeneous devices, such knowledge reuse can also significantly reduce the network-level map construction and maintenance costs compared with existing CKMs. Moreover, the standardized and vendor-neutral knowledge representation of uCKM can provide a common knowledge interface for heterogeneous devices and vendors, which facilitates the cross-vendor coordination and improves interoperability among diverse network entities.
 
 \subsubsection{Cross-Task Channel Knowledge Transfer} 
 Beyond facilitating cross-device channel knowledge transfer, uCKM allows diverse downstream tasks to reuse a common knowledge prior. Specifically, although communication, localization, sensing, and navigation tasks have different objectives, they extract the task-specific information from the same wireless propagation environment. Accordingly, each task can obtain its required information based on uCKM on demand through appropriate processing. As a result, uCKM not only reduces redundant task-specific map construction and maintenance, but also improves the utilization efficiency of wireless propagation knowledge, thus advancing the development of environment-aware applications. More broadly, by serving as a common environment prior base, uCKM provides a foundation for the wireless digital twin and AI-native environment-aware networks.

 \subsection{All for uCKM}
 
 \subsubsection{Cross-Device Collaborative Knowledge Acquisition}
 In contrast to existing CKMs that are typically constructed based on the isolated measurements, uCKM fuses globally aligned observations crowdsourced from heterogeneous devices. Such a mechanism improves the construction of uCKM in terms of coverage and reliability. On one hand, crowdsourced observations from heterogeneous devices can progressively fill the spatial blind spots that are difficult to be covered via the isolated measurement, such as indoor corridors, urban corners, low-altitude regions, and rarely visited regions. As the observations accumulate, uCKM can obtain a more comprehensive representation of the wireless propagation environment. On the other hand, observations from multiple independent devices can be exploited for consensus-based knowledge validation before incorporating them into uCKM. Specifically, consensus should be made based on observations with spatial and temporary correlations, and fused with reliability-based weights. Then, consistent deviations reported by multiple reliable and independent devices may indicate environmental changes, whereas isolated observations inconsistent with those from other reliable devices are more likely to be caused by device-specific anomalies. 
 Thus, multi-source consensus can support anomaly detection and trustworthy knowledge fusion, thus enhancing the reliability of uCKM construction. 
 
 %For instance, consistent deviations reported by multiple independent devices in the same region may indicate that the environment has changed. On the contrary, if only one device reports an abnormal observation, such an abnormality is more likely to be caused by the device itself. Thus, multi-source consensus can support anomaly detection and trustworthy knowledge fusion, thus enhancing the reliability of uCKM construction. 
 
 \subsubsection{Cross-Task Complementary Knowledge Enrichment}
 Besides the cross-device collaborative knowledge acquisition, diverse downstream tasks contribute complementary observations that broaden and refine the knowledge captured by the uCKM. Specifically, communication tasks provide the CSI, beam management information, and link quality variations; localization tasks contribute the location-tagged multi-path fingerprints; and sensing tasks offer the information of clutters, scatterers, obstacles, and environmental changes. Instead of being the redundant replicas, these complementary knowledge jointly yield a richer, fine-grained, and more physically interpretable representation of the underlying propagation environment for uCKM. Moreover, by exploiting the real-time task feedbacks, uCKM can identify the environmental changes and update the outdated channel knowledge. Thus, cross-task participation not only enriches the knowledge dimension of uCKM, but also maintains its freshness via the closed-loop refinement. 
 
 % >>>>>>>>>>>>>SECTIONS III -  here >>>>>>>>>>>> 
\section{Challenges in uCKM and Potential Solutions}
 Despite the above appealing benefits, uCKM is also faced with several new challenges. 
 
 \subsection{Data Acquisition and Dataset Construction}
 The foundational prerequisite for constructing a high-fidelity uCKM is the acquisition of massive high-quality measurement data. Traditionally, high-fidelity radio measurements are mainly obtained through dedicated channel sounding, which can provide detailed channel characteristics. However, the high measurement overhead renders it difficult to support the dense uCKM deployment. To address this limitation, the real-world measurement data can be complemented by multi-modal sensing data, such as radar, light detection and ranging (LiDAR), and vision. By capturing the environmental geometry including reflectors, scatterers, obstacles, mobile objects, and semantic scene information, the multi-modal data can provide the physical context behind radio measurements, which is particularly beneficial for the uCKM construction that relies on multi-path parameters. Meanwhile, simulation-generated data can serve as a scalable supplement to real-world measurement data \cite{li2025digital}. To improve the reliability of the simulation data, the simulator calibration and bias correction have been proposed to effectively mitigate the simulation-to-reality gap.

 In addition, constructing a reliable dataset remains non-trivial, since it is not merely to accumulate the data, but to integrate the real-world measurement and multi-modal sensing data into a spatially consistent and location-specific data repository. To achieve this goal, the collected data should be arranged in a unified data format with metadata such as location, timestamp, and source type. Moreover, real-world measurement, multi-modal, and simulation data should be spatially and temporally aligned in a common global coordinate system, thus enabling a quality-aware dataset construction for uCKM.
 
 \subsection{Device-Decoupled Path Knowledge Extraction}
 The collected raw data are in fact device-filtered observations and inevitably coupled with the device-specific transceiver configurations. Thus, a key challenge is to accurately decouple the universal propagation knowledge from the heterogeneous observations, which may suffer from device-dependent errors. First, accurately characterizing transceiver characteristics, including antenna pattern, hardware impairments, and unknown calibration state, is a non-trivial task, since these factors can be time-varying or even proprietary. Second, even if the transceiver characteristics can be perfectly characterized and removed, the intrinsic resolution differences among devices will lead to different observed multi-path representations. Specifically, the delay and angular resolutions are determined by the system bandwidth and physical array aperture, respectively, and low-resolution devices may be unable to resolve closely spaced propagation paths and observe them as a single equivalent path. This resolution mismatch renders the fine-grained path recovery an ill-posed inverse problem. 
 
 To address these challenges, device-decoupled path knowledge extraction should jointly account for transceiver characteristics, device resolution, and propagation physics. Specifically, the transceiver characteristics can be obtained based on the offline calibration, pilot-aided estimation, and auxiliary sensors. Meanwhile, path knowledge extraction should take into account the intrinsic resolution of each device. In this context, integrated sensing and communication (ISAC) is an effective approach for extracting path knowledge, by leveraging its sensing capability to estimate the geometric and propagation properties of multi-path components. The resulting information provides physics-informed prior for separating closely spaced paths, resolving ambiguities caused by limited device resolution, and associating multi-path components across heterogeneous devices. Nevertheless, the remaining estimation errors should be explicitly considered during the global coordinate transformation and subsequent knowledge fusion. Furthermore, physics-informed AI models can fuse heterogeneous observations, associate paths across devices, and quantify uncertainties, thus transforming the device-specific observations into the fundamental and reusable channel knowledge.

 \subsection{Discrete-to-Continuous uCKM Construction in Dynamic Environments}
 After obtaining the universal propagation knowledge at sampled locations, constructing a continuous uCKM from discrete and sparse observations remains challenging. Unlike a standard interpolation problem, uCKM construction needs to capture the abrupt spatial transitions in multi-path characteristics, e.g., the dominant LoS path may suddenly vanish and be replaced by NLoS reflections at a street corner. In such cases, simply interpolating neighboring observations may yield physically inconsistent channel knowledge \cite{wang2026toward}. Moreover, training a deep neural network (DNN) to map the location to channel knowledge may also over-smooth the predicted channel knowledge near propagation boundaries, since neural networks tend to favor smooth function approximations. To tackle this issue, the DNN should be designed in an environment-aware and physics-informed manner, by embedding physical knowledge and constraints into data representations, loss functions, or network architectures. Besides, the DNN can be trained to jointly predict the multi-path visibility and parameters, where the multi-path visibility labels can be obtained from ray-tracing simulations and measurements. Such a design can explicitly capture the path birth-death behaviour and alleviate over-smoothing near propagation boundaries. Furthermore, the discrete path visibility and continuous multi-path parameters, such as angles and delays, can be jointly represented by modeling the overall pdf as a weighted combination of the conditional distributions associated with different path-visibility states. The performance evaluation of such construction is worthy of more in-depth studies in the future.

 \begin{figure}
 \setlength{\abovecaptionskip}{-0.05cm}
 \setlength{\belowcaptionskip}{-0.1cm}
 \centering
 \subfigure[]{
 \begin{minipage}[t]{0.5\textwidth}
 \centering
 \centerline{\includegraphics[width=2.0in,height=1.6in]{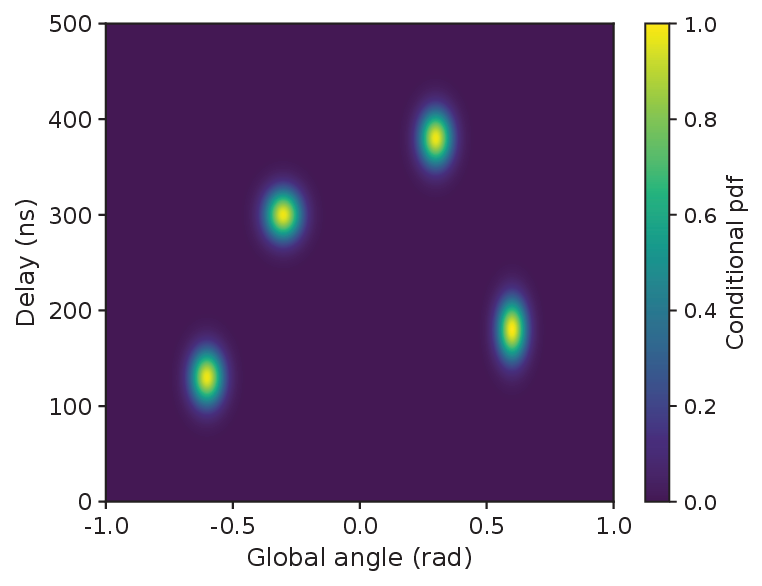}}
 \end{minipage}
 }
 \subfigure[]{
 \begin{minipage}[t]{0.5\textwidth}
 \centering
 \centerline{\includegraphics[width=2.0in,height=1.6in]{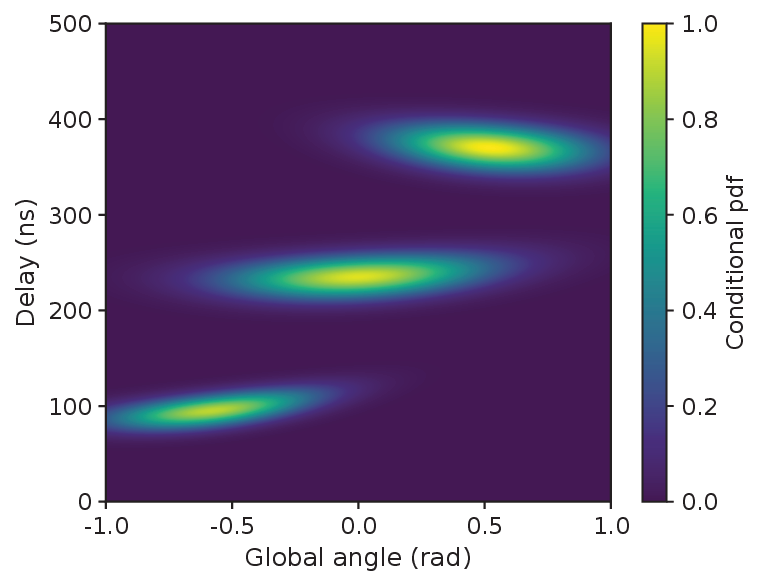}}
 \end{minipage}
 }
 \caption{Illustration of uCKM for two locations, where the location-specific channel knowledge at each location, including the angle and delay represented in the global coordinate system, is represented by a corresponding conditional pdf: a) location 1; b) location 2.}
 \label{fig:locationConditionalPdf}
 \vspace{-0.5cm}
 \end{figure}
 %[width=3.0in,height=2.4in]
 
 On the other hand, representing the mapping from the transceiver location to channel knowledge as a conditional pdf for uCKM provides a feasible solution to handle environmental dynamics, which characterizes the joint distribution of the channel knowledge vector given the location. As illustrated in Fig.~\ref{fig:locationConditionalPdf}, the two locations exhibit distinct probabilistic distributions, which not only provide the most probable realization of the location-specific channel knowledge, but also quantify the associated uncertainty, e.g., through the variance. However, a key challenge lies in accurately learning the ground-truth conditional pdf. Model-based pdf representations, such as Gaussian, Gaussian mixture, and Rayleigh distributions, are compact and interpretable, while they may fail to accurately approximate the ground-truth conditional pdf in highly dynamic and complex propagation environments. To this end, model-free approach can be adopted, and generative AI is particularly suitable for such an approach, since it can train an AI model to approximate the ground-truth distribution and generate the new channel knowledge realizations according to the learned distribution \cite{fu2026ckmdiff,zhao20253d}.

 \subsection{Privacy-Preserving and Trustworthy uCKM Deployment}
 As uCKM provides wireless environment prior for heterogeneous devices and diverse tasks, its large-scale deployment introduces challenges in privacy preservation and trustworthiness. On one hand, the reported channel knowledge may unintentionally expose the user privacy, since the embedded multi-path fingerprints can be exploited by malicious users to infer the locations and trajectories of target users. Besides, legitimate users may expose sensitive mobility patterns when querying uCKM, as a sequence of location-specific queries can be linked to infer their destinations, frequently visited areas, or locations of interest. Moreover, the channel knowledge accessed by malicious users can be exploited to launch the targeted eavesdropping, jamming, or spoofing attacks. On the other hand, in crowdsourced data acquisition scenarios, the malicious users can inject the forged or incorrect measurement data to poison the map, thus misleading the downstream tasks.
  
 To mitigate the above issues, security mechanisms should be integrated across data acquisition, access, and utilization of uCKM. First, privacy-preserving techniques, such as differential privacy and homomorphic encryption, can be applied to allow devices to collaboratively construct and update uCKM without exposing raw data. Meanwhile, multi-source validation and anomaly detection are needed to identify the forged or incorrect measurement data, so as to mitigate the map contamination. Second, during the access process, authentication and access control are essential to prevent the leakage of sensitive propagation knowledge. Specifically, the identities of querying entities should be verified before granting access, and different devices may be provided with channel knowledge at different granularities based on their roles and trust levels. Last, devices and downstream tasks should explicitly account for the uncertainty and reliability of uCKM, thus establishing robust decision-making mechanisms when uCKM is potentially contaminated. 
  
 % >>>>>>>>>>>>>SECTIONS IV -  here >>>>>>>>>>>> 
  \vspace{-0.3cm}
 \section{Simulation Results} 
\begin{figure*}[!t]
\centering
\subfigure[]{%
\begin{minipage}[t]{0.31\textwidth}
    \centering
    \includegraphics[width=\linewidth]
    {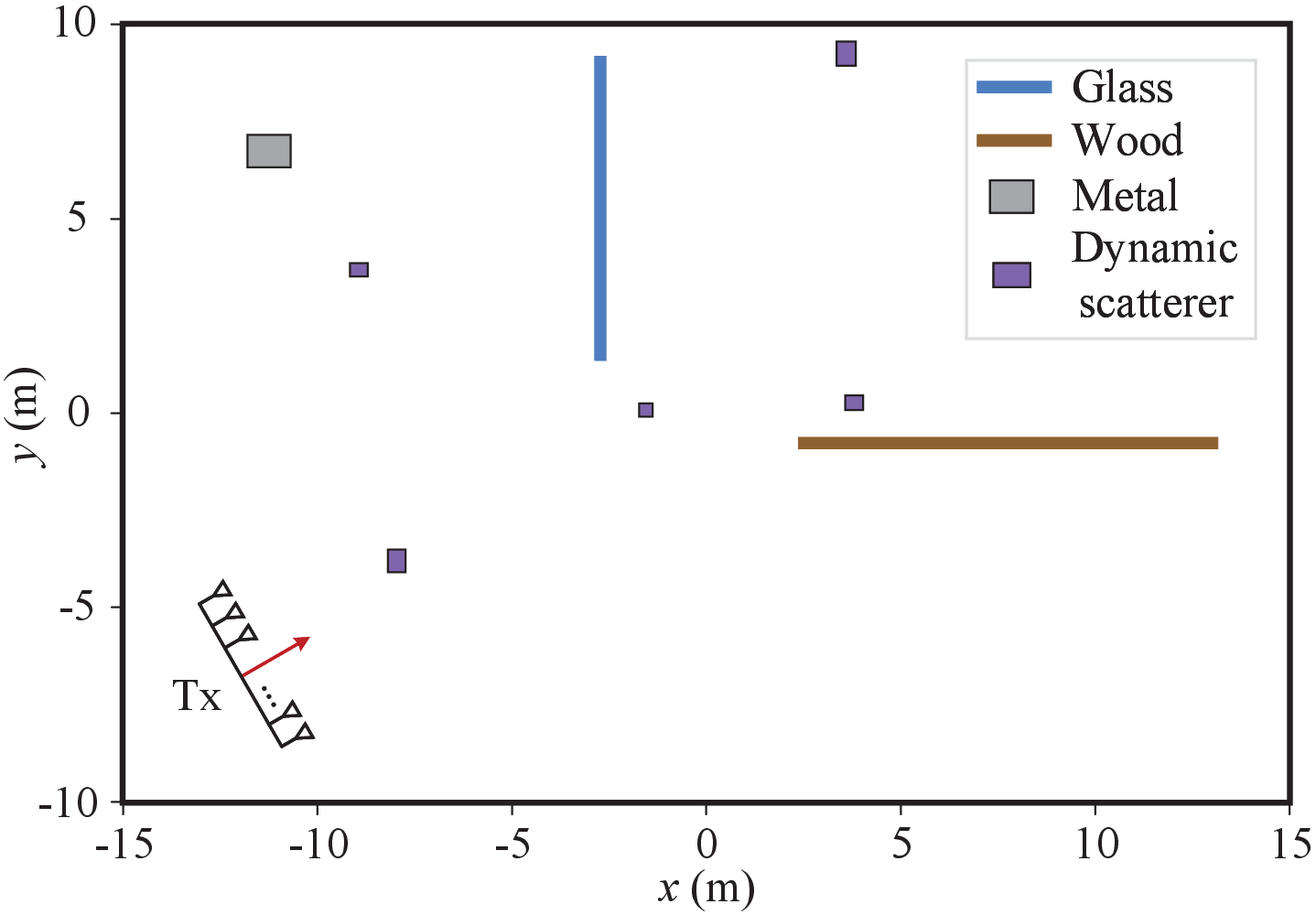}
\end{minipage}%
}
\hfill
\subfigure[]{%
\begin{minipage}[t]{0.31\textwidth}
    \centering
    \includegraphics[width=\linewidth]
    {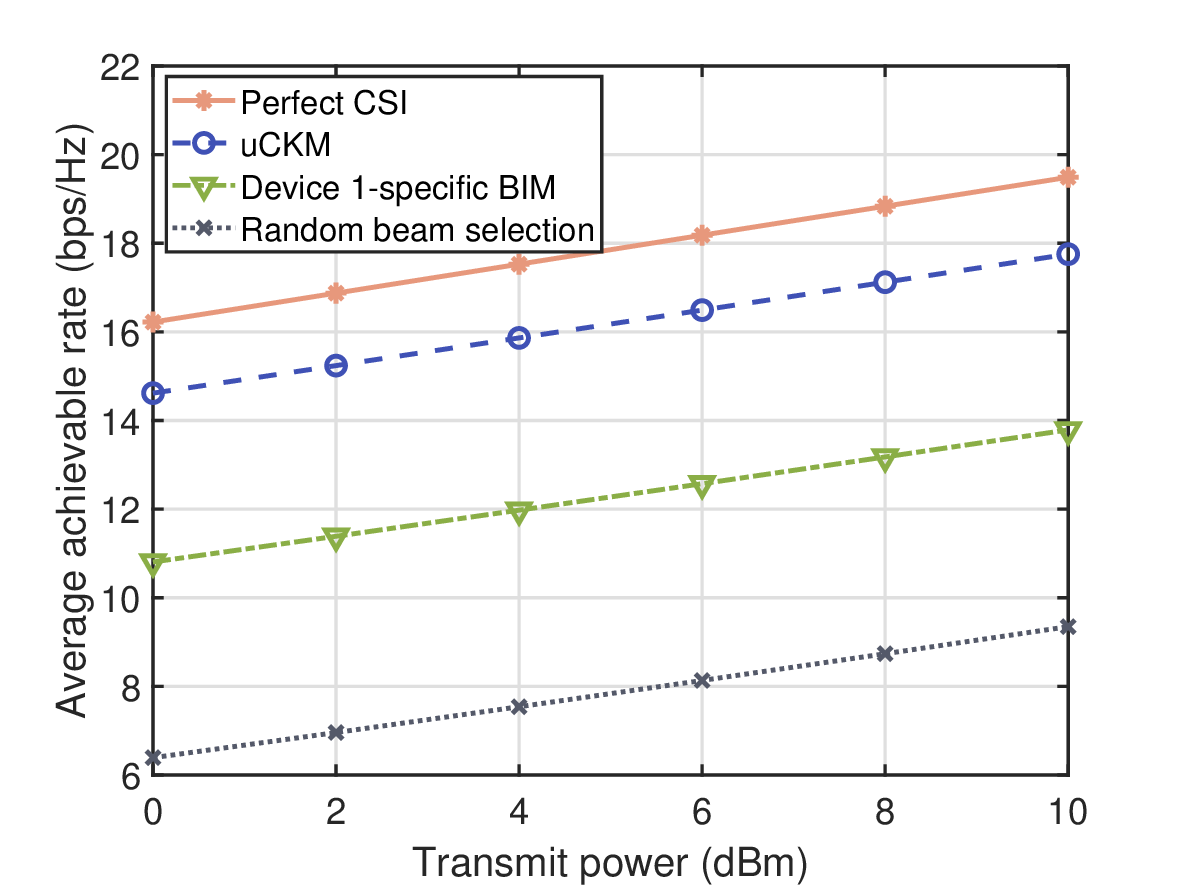}
\end{minipage}%
}
\hfill
\subfigure[]{%
\begin{minipage}[t]{0.31\textwidth}
    \centering
    \includegraphics[width=\linewidth]
    {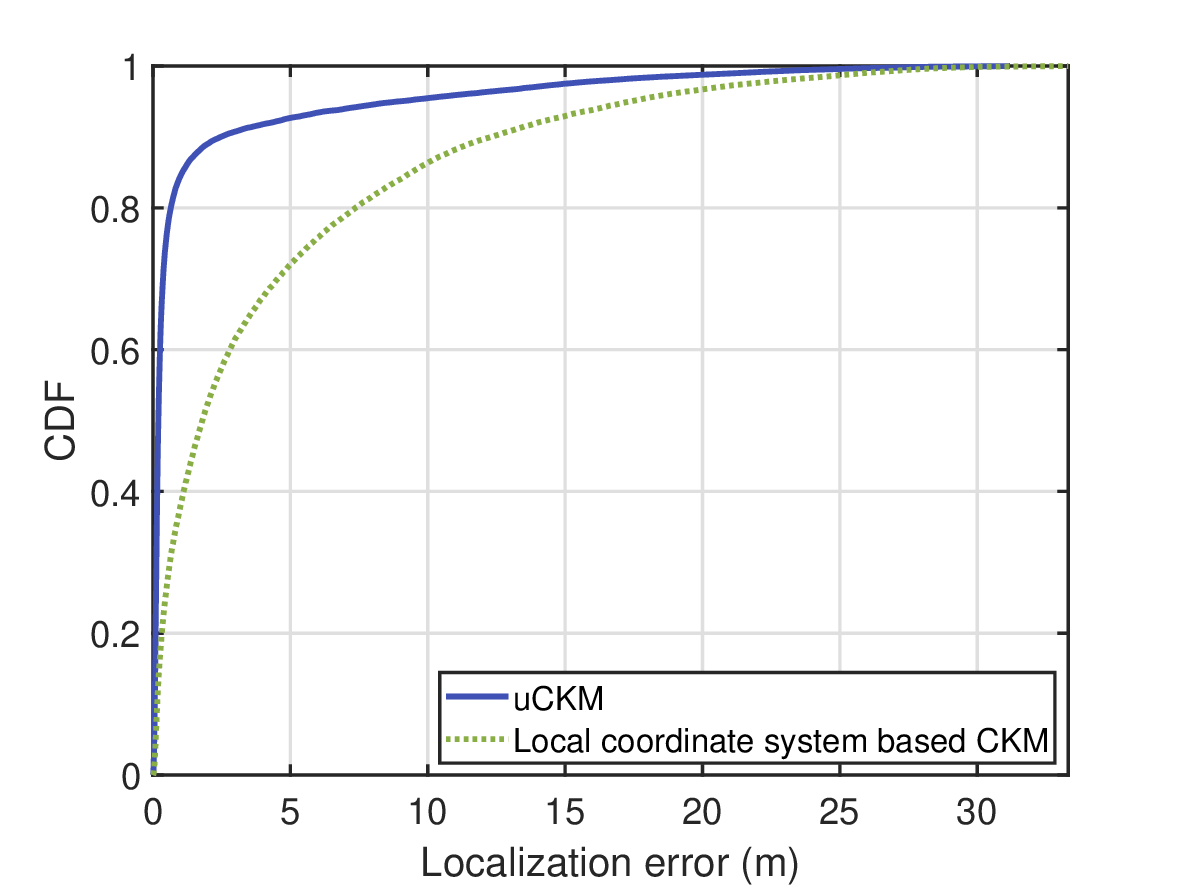}
\end{minipage}%
}
\caption{Performance evaluation of uCKM for communication and localization: a) simulation setup, where the dynamic scatterers correspond to one realization, while their locations vary across different sub-scenes; b) average achievable rate versus the transmit power; c) the CDF of the localization error.}
\label{fig:simulationResultuCKM}
\end{figure*}

 In this section, simulation results are presented to evaluate the cross-device and cross-task transferability of uCKM. The open-source ray-tracing tool Sionna is employed to simulate the wireless propagation characteristics and generate the corresponding multi-path parameters. The simulation setup is shown in Fig.~\ref{fig:simulationResultuCKM}(a). The carrier frequency is $3.5$ GHz, the Tx is equipped with a uniform linear array (ULA) with $N_t = 64$ antennas, where adjacent antennas are separated by half wavelength. We consider $K =3$ heterogeneous devices, each equipped with a ULA, where the number of receive antennas are $N_r = 16$, $8$, $4$, respectively. The configurations may correspond to an autonomous vehicle, a robot, and a smartphone, respectively. The device orientations are varied by adjusting their yaw angles, and the devices follow a local random-walk sampling strategy within a rectangular region, with its length and width given by $30$ m and $20$ m, respectively. A total of $64$ sub-scenes are generated with the same static scatterers shown in Fig.~\ref{fig:simulationResultuCKM}(a), but different realizations of five randomly located dynamic scatterers. Moreover, the region is discretized into multiple grid cells of size $0.25$ m $\times$ $0.25$ m.

% \begin{figure}[!t]
% \centering
% \centerline{\includegraphics[width=3.5in,height=2.75in]{crossDeviceRateVersusSNR.eps}}
% \caption{Average achievable rate versus SNR.}
% \label{fig:crossDeviceRateVersusSNR}
% \end{figure}  
  
 Fig.~\ref{fig:simulationResultuCKM}(b) shows the average achievable rate over the three devices versus the transmit power. For comparison, the following benchmark schemes are considered: 1) Perfect CSI: The device-specific CSI of each device is perfectly available, and the optimal transmit and receive beam pair is then selected from the predefined codebooks via exhaustive beam search, which provides an upper bound on the achievable rate; 2) Device 1-specific BIM: The BIM constructed for device 1 is directly reused across heterogeneous devices due to the lack of knowledge of other devices' configurations; 3) Random beam selection: five beam pairs are randomly selected, and the beam pair with the highest achievable rate is chosen. Moreover, uCKM learns the location-conditional pdf of the global AoA, AoD and propagation delay using a Gaussian mixture distribution, which stores the corresponding weights, mean vectors and covariance matrices. For beam pair selection, the angular samples are generated based on the mean vector and covariance for each Gaussian mode, and ranked by their mixture-weighted scores. Subsequently, the sampled global angles are transformed into the local coordinate systems according to the yaw angles of transceivers, and mapped to the nearest entries of the corresponding discrete Fourier transform (DFT) codebooks. After removing the duplicate beam pairs, the five beam pairs for each device are selected and the beam pair with the highest achievable rate is chosen.
 
 It is observed that the proposed uCKM scheme achieves a performance close to that of the perfect CSI scheme, thanks to the flexible transferability of channel knowledge across heterogeneous devices. Besides, the slight performance gap is mainly due to the mismatch between the accurate CSI and the location-conditional channel knowledge distribution represented by uCKM. It is also observed that the proposed uCKM scheme significantly outperforms the benchmark schemes of device-1-specific BIM and random beam selection. This is because device 1-specific BIM scheme is highly entangled with the antenna orientation of device 1 and thus cannot be directly transferred to the other devices, while uCKM stores the underlying wireless propagation knowledge and can adapt it to the device-specific beam domain. 
 
 Fig.~\ref{fig:simulationResultuCKM}(c) shows the cumulative distribution function (CDF) of localization error, where the uCKM employed for the above communication task is directly utilized for localization. Specifically, given the device observations, uCKM evaluates the posterior probability scores of candidate cells and infers the most probable device location from the grid cell with the maximum score. As a comparison, the benchmark scheme of CKM constructed in the local coordinate systems of devices is considered. It is observed that uCKM yields a better localization performance than the local coordinate system based CKM. This is expected since the channel knowledge is represented in the unified global coordinate system for uCKM, and the result demonstrates its cross-task transferability. By contrast, for the local coordinate system representation, the wireless propagation environment may correspond to different features, thus leading to the knowledge mismatch and degraded localization accuracy. 

 %%%% >>>>>>>>>>>>>SECTIONS V -  here >>>>>>>>>>>>

 \section{Future Directions}
 In the future, there are several important directions worthy of in-depth studies. 
 
 \subsection{Data-Efficient Construction of uCKM}
 The prediction accuracy of uCKM depends on the amount of data, and a larger spatial sample density generally improves the prediction accuracy by increasing the likelihood of finding nearby observations with high spatial correlations \cite{xu2024much}. However, indiscriminately increasing the spatial sample density may result in the excessive measurement overhead with marginal performance improvement. Thus, an important direction is to develop data-efficient strategies that jointly consider the sampling density and measurement locations for uCKM construction, so as to achieve the reliable prediction with affordable cost.
 
 \subsection{World Model-Based uCKM Update}
 World models may serve as a promising direction for proactive uCKM updates in dynamic environments. Instead of only learning the mapping from the transceiver location to channel knowledge, a wireless world model aims to capture the latent state and temporal evolution of the radio environment, thus guiding the proactive uCKM updates before significant prediction errors occur. 
 
 \subsection{Standardization and Commercial Deployment of uCKM}
 Despite the technical potential of uCKM, its scalable deployment in future wireless networks requires further studies. On one hand, the common knowledge representations, interface protocols, and map update mechanisms should be standardized to enable cross-vendor interoperability. On the other hand, the commercial deployment involves the operation and maintenance mechanism, and business ecosystem design, so as to promote uCKM as an economically viable and operationally sustainable wireless environment prior.

 %%%% >>>>>>>>>>>>>SECTIONS VI -  here >>>>>>>>>>>>
 \section{Conclusion}
 This paper presented an overview of uCKM as a shared wireless environment prior base for environment-aware wireless networks, where uCKM aims to enable channel knowledge transfer across heterogeneous devices and diverse downstream tasks. We elaborated the benefits of uCKM through two complementary visions, termed ``uCKM for All'' and ``All for uCKM'', and discussed the challenges for realizing uCKM and potential solutions. Simulation results demonstrated the feasibility of cross-device and cross-task knowledge transfer enabled by uCKM. %We hope that the discussed research directions would stimulate the further investigation into uCKM in the future.

%\begin{appendices}
%\section{}
%\end{appendices}

%\ifCLASSOPTIONcaptionsoff
%  \newpage
%\fi

\bibliographystyle{IEEEtran}
\bibliography{refUCKM}

\end{document}